\documentclass[conference]{IEEEtran}
\usepackage{cite}
\ifCLASSINFOpdf
   \usepackage[pdftex]{graphicx}
\else

\fi

\usepackage{amsmath}
\usepackage{slashed}
\usepackage{array}
\usepackage{amsmath,amssymb,amsthm,verbatim,amsfonts,amscd,graphicx,float,physics}

\begin{document}

\title{Thermo field dynamics on a quantum computer}

\author{\IEEEauthorblockN{Raffaele Miceli}
\IEEEauthorblockA{Physics Department\\
Stony Brook Univerity\\
Stony Brook, New York 11794\\
Email: raffaele.miceli.32@gmail.com}
\and
\IEEEauthorblockN{Michael McGuigan}
\IEEEauthorblockA{Computational Science Initiative\\
Brookhaven National Laboratory\\
Upton, New York 11973\\
Email: mcguigan@bnl.gov}}

\maketitle

\begin{abstract}
In this project we develop a quantum algorithm to realize finite temperature simulation on a quantum computer. As quantum computers use real-time evolution we did not use the imaginary time methods popular on classical algorithms. Instead, we implemented a real-time therom field dynamics formalism, which has the added benefit of being able to compute quantities that are both time- and temperature-dependent. To implement thermo field dynamics we apply a unitary transformation \cite{Das:1989cj} to discrete quantum mechanical operators to make new Hamiltonians with encoded temperature dependence. The method works normally for fermions, which have a finite representation, but needs some modification to work with bosons. These Hamiltonians are then processed into a Pauli matrix representation in order to be used as input for IBM's Qiskit package \cite{Qiskit}. We then use IBM's quantum simulator to calculate an approximation to the Hamiltonaian's ground state energy via the variational quantum eigensolver (VQE) algorithm \cite{Peruzzo:2013nat}. This approximation is then compared to a classically calculated value for the exact energy. The thermo field dynamics quantum algorithm has general applications to material science, high-energy physics and nuclear physics, particularly in those situations involving real-time evolution at high temperature.
\end{abstract}

\section{Introduction - Discrete Quantum Mechanics}

To work on quantum computers, we need to discretize \cite{Jagannathan:1981rh} the operators of regular quantum mechanics. When we create these operators, we need to choose a basis in which to work. In our projects we focus on two of these bases, the position basis and the energy basis.

To create our operators for the position basis, we start by defining a lattice:

\begin{equation}
    \ell(a,n) = \frac{2a-1-n}{2}
\end{equation}

The index $a$ runs from $1$ to $2^n$, where n is the number of qubits we want to use. ($2^n$ is the number of lattice sites.) Using this lattice we define the position operator X:

\begin{equation}
    \bra{j} X_{pos} \ket{k} = \sqrt{\frac{2\pi}{n}} \ \ell(j,n) \ \delta(j,k)
\end{equation}

To define the conjugate momentum operator, we use the discrete Fourier transform:

\begin{equation}
    \bra{j} F_n \ket{k} = \frac{e^{\frac{2\pi i}{n}\ell(j,n)\ell(k,n)}}{\sqrt{n}}
\end{equation}

The momentum operator $P$ is then:

\begin{equation}
    P_{pos} = F^\dag X_{pos} F
\end{equation}

In the energy basis, we start by defining the annihilation operator for the discrete harmonic oscillator:

\begin{equation}
    \bra{j} A \ket{k} = \sqrt{j} \ \delta(j,k-1)
\end{equation}

We then define the energy basis position and momentum operators as follows:

\begin{equation}
    X_{en} = \frac{A + A^\dag}{\sqrt{2}} \ , \  P_{en} = \frac{i(A - A^\dag)}{\sqrt{2}}
\end{equation}

These operator relations can be inverted to obtain the creation and annihilation operators in the position basis. We compare the spectra of the bosonic harmonic oscillator Hamiltonian in both bases below:

\begin{equation}
    H_P = \frac{X^2 + P^2}{2} \ , \ H_E = A^\dag A + I/2
\end{equation}

Here $I$ is the identity matrix. Below we compare the spectra (eigenvalues) of these two Hamiltonians. We can observe from the plots that they match almost exactly in about half of the values.

\begin{figure}[h]
    \centering
     \includegraphics[width=0.8\linewidth]{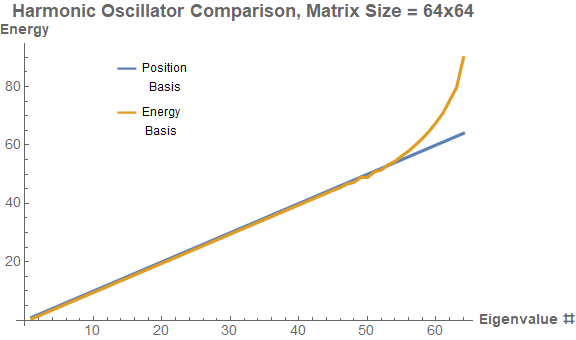}
    \caption{Spectrum of the harmonic oscillator Hamiltonian, using both position and energy bases.}
    \label{fig:compare}
\end{figure}

In the position basis, the eigenvectors of our operators represent the probability that a particle will be found at a particular position on our lattice. In the energy basis, they represent the probability that a particle will be in a particular harmonic oscillator energy state. We can recover the position basis state by multiplying the components of the eigenvector by the corresponding Hermite function $H_n(x)$ and taking a sum:

\begin{equation}
    \psi(x) = \sum\nolimits_i \omega_i H_i(x)
\end{equation}

\begin{equation}
    H_n(x) = \frac{e^{-x^2/2}}{\sqrt{2^n n! \sqrt{pi}}} \bigg( x - \frac{d}{dx} \bigg)^n \cdot 1
\end{equation}

Discrete quantum mechanics is also developed by Singh in \cite{Singh:2018qzk}, by Atakishiyev in \cite{Atakishiyev:2008}, by Lorente in \cite{Lorente:2001}, and by Barker in \cite{Barker:2000}.

\section{Thermo Field Dynamics}

Thermo field dynamics introduces temperature to a previously defined system through what's called a Bogoliubov transformation \cite{Bogoljubov:1958}. This is an (ideally) unitary operator which can be used on other operators as well as states. The theory was developed by Takahashi and Umezawa in \cite{Takahashi:1996zn} and by Das in \cite{Das:1989cj}. It is also covered by Cottrell in \cite{Cottrell:2018ash} and by Wu in \cite{Wu:2018nrn}.

We will focus on a single fermion system for this explanation. We start by creating a thermo double particle for our fermion. We label the real fermion's raising operator $c_L$ and the thermo double's $c_R$. We construct the elevated operators from the single fermion operator via the Kronecker product:

\begin{equation}
    c_L = c \otimes I, \ c_R = \sigma_z \otimes c
\end{equation}

The Pauli matrix $\sigma_z$ is used in the construction of $c_R$ in order to maintain the anti-symmetry between fermions. We then use these operators to create a new operator:

\begin{equation}
    G = -i \theta(\beta)(c_R c_L - c_L^\dag c_R^\dag)
\end{equation}

Here $\theta$ is a factor that we use to introduce temperature dependence:

\begin{equation}
    \tan\theta (\beta) = e^{-\beta m/2}, \ \beta = 1/kT
\end{equation}

With this operator, or rather its exponentiation, we can transform an arbitrary zero-temperature operator $A(0)$ into a temperature-dependent one:

\begin{equation}
    A(\beta) = e^{iG}A(0)e^{-iG}
    \label{eq:ftemp_op}
\end{equation}

We can also use G to create temperature-dependent states from zero-temperature ones:

\begin{equation}
    \ket{0(\beta)} = e^{-iG}\ket{0}
    \label{eq:ftemp_state}
\end{equation}

To calculate the expectation value of a finite-temperature operator in a zero-temperature state - or equivalently the expectation value of a zero-temperature operator in a finite-temperature state - we use the following expression:

\begin{equation}
    E_0(\beta) = \langle 0|e^{iG} H e^{-iG} |0 \rangle
\end{equation}

Here $H$ is simply the Hamiltonian for a single fermion created from our $c_L$ operator, not including the zero-point energy, with $m$ being the mass of the fermion:

\begin{equation}
    H = m(c_L^\dag c_L)
\end{equation}

\section{Discrete Thermo Field Dynamics}

Because the fermion is can be represented discretely, the previous construction of G can be used. In the case of the boson, which cannot be represented with finite matrices, we need to use a different construction based on the maximally entangled state:

\begin{equation}
    e^{-iG} =  \frac{\sum\nolimits_{j=0}^{n} \big( e^{-m\beta a_R^\dag a_L^\dag/2} \big)^j/j!}{Z(\beta)^{1/2}}
\end{equation}

where $Z(\beta)$ is the partition function:

\begin{equation}
    Z(\beta) = \sum\nolimits_{n} e^{-\beta E_n}
\end{equation}

In the case of the fermion we can use a simple equation for the exact energy:

\begin{equation}
    E_{0E,F}(\beta) = \frac{e^{-\beta m}}{1+e^{-\beta m}}
    \label{fig:E0EF}
\end{equation}

While for the boson we need to explicitly sum over the finite number of energy states:

\begin{equation}
    E_{0E,B}(\beta) = \frac{\sum_{j=0}^{n} e^{- \beta j m} j m}{\sum_{j=0}^{n} e^{- \beta j m}}
    \label{fig:E0EB}
\end{equation}

To visualize the thermal ground state wavefunction in the position basis, we start with it in the energy basis:

\begin{equation}
    \ket{0(\beta)} = \frac{\sum\nolimits_n e^{-\beta E_n/2} \ket{n,\Tilde{n}}}{Z(\beta)^{1/2}}
\end{equation}

Then we contract it with the vacuum position basis ground state, which are the Hermite functions $H_n$ from before:

\begin{equation}
    \begin{split}
        \braket{x,\Tilde{x}}{0(\beta)} = \frac{\sum\nolimits_n \braket{x,\Tilde{x}}{n,\Tilde{n}} e^{-\beta E_n/2}}{Z(\beta)^{1/2}} \\
        = \frac{\sum\nolimits_n H_n(x) H_n(\Tilde{x}) e^{-\beta E_n/2}}{Z(\beta)^{1/2}} \\
        = 
    \end{split}
\end{equation}

Below we plot this wavefunction for various values of $\beta$:

\begin{figure}[H]
    \centering
    \includegraphics[width=0.75\linewidth]{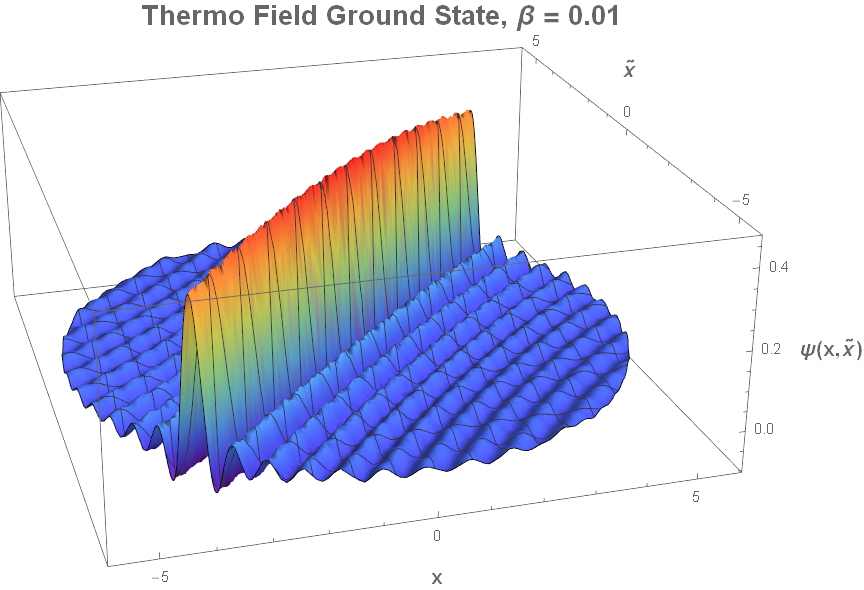}
    \caption{Position ground state wavefunction with  $\beta$ = 0.01.}
    \label{fig:gs_b001}
\end{figure}

\begin{figure}[H]
    \centering
    \includegraphics[width=0.75\linewidth]{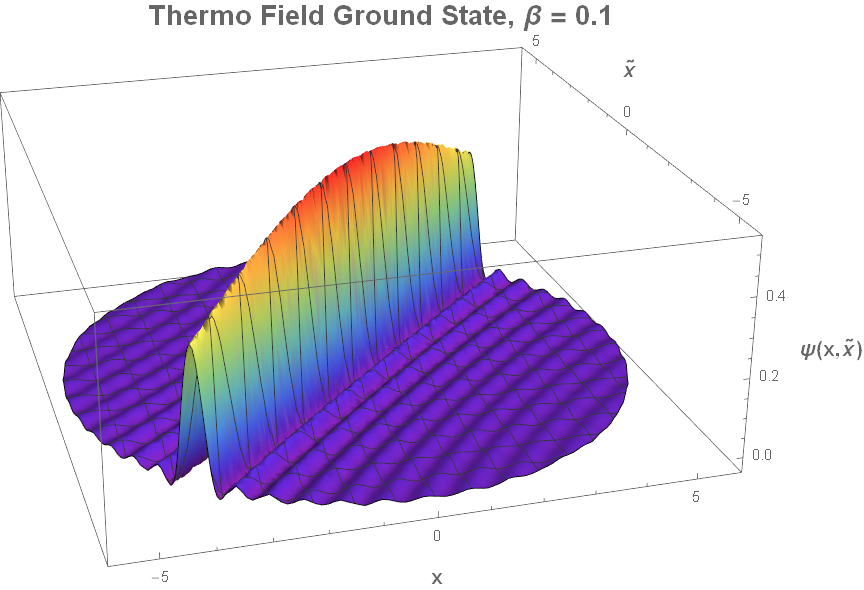}
    \caption{Position ground state wavefunction with  $\beta$ = 0.1.}
    \label{fig:gs_b01}
\end{figure}

\begin{figure}[H]
    \centering
    \includegraphics[width=0.75\linewidth]{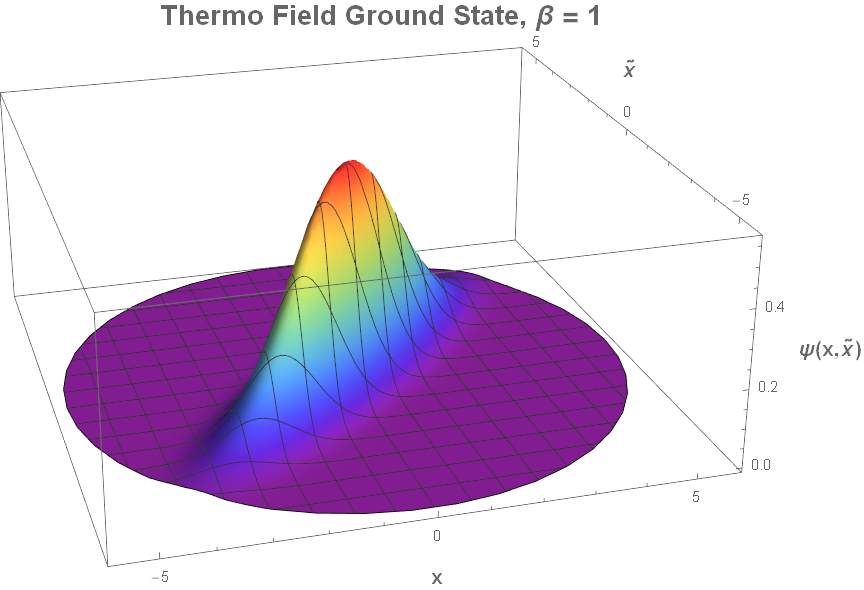}
    \caption{Position ground state wavefunction with  $\beta$ = 1.}
    \label{fig:gs_b1}
\end{figure}

\begin{figure}[H]
    \centering
    \includegraphics[width=0.75\linewidth]{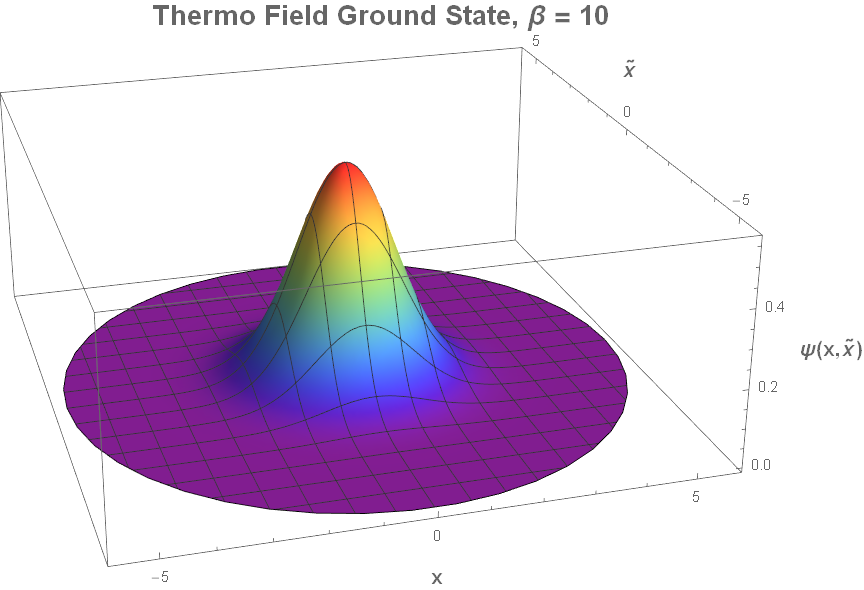}
    \caption{Position ground state wavefunction with  $\beta$ = 10.}
    \label{fig:gs_b10}
\end{figure}

We can see that as temperature decreases ($\beta$ increases), the ground state moves closer and closer to the product of two Gaussians, as we'd expect for the vacuum ground state.

\section{Quantum Computing Methods}

If we naively attempt to run the variational quantum eigensolver (VQE) \cite{Peruzzo:2013nat} algorithm on a finite-temperature Hamiltonian, using the procedure described in \cite{Miceli:2018} we get something like this:

\begin{figure}[H]
    \centering
    \includegraphics[width=0.9\linewidth]{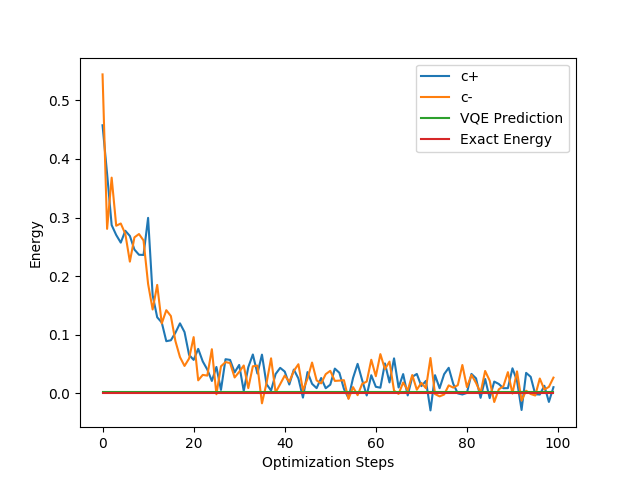}
    \caption{Convergence plot from a run on a fermionic harmonic oscillator Hamiltonian modified for finite temperature.}
    \label{fig:ftemp_plot}
\end{figure}



The process seems to converge to something, but it does not match the ground state energy we get by direct calculation using equation \ref{fig:E0EF}. This is because the Bogoliubov transformation we use to implement thermo field dynamics is in essence a change of basis, which doesn't affect the eigenvalues of our Hamiltonian matrix. Since the VQE algorithm simply looks for the lowest eigenvalue of the matrix, it is not sensitive to this modification. This means we need to explore thermo field dynamics in quantum computing using a different method.

If we take another look at equation \ref{eq:ftemp_state}, we can observe that the action of the Bogoliubov transformation on the vacuum ground state is identical to that of a time-evolution operator, only that here time is replaced by inverse temperature. So if we can figure out how to represent the Bogoliubov transformation in terms of quantum gates, we could set up a quantum circuit to observe its action on the vacuum ground state directly.

The Qiskit Python package \cite{Qiskit} has a function which can decompose a given unitary matrix into quantum gates, but it currently only works for 2-qubit ($4\times4$) unitary matrices. This limits us to simulating thermo field dynamics on a single fermion, since we need to leave room for the thermo double.

\begin{figure}[H]
    \centering
    \includegraphics[width=0.9\linewidth]{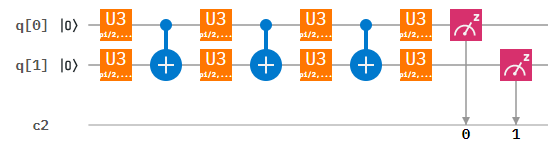}
    \caption{Quantum circuit for thermo field operator as created by Qiskit.}
    \label{fig:2qwalk_circui}
\end{figure}

The Qiskit script we wrote starts with two qubits in the vacuum state. Then the operator $e^{-iG(\beta)}$, converted into quantum gates, acts on the qubits. The qubits are then measured and the state is recorded. This process is repeated many times for each value of $\beta$ we want to probe, and finally the results are plotted.

\section{Quantum Computing Results}

\begin{figure}[H]
    \centering
    \includegraphics[width=0.8\linewidth]{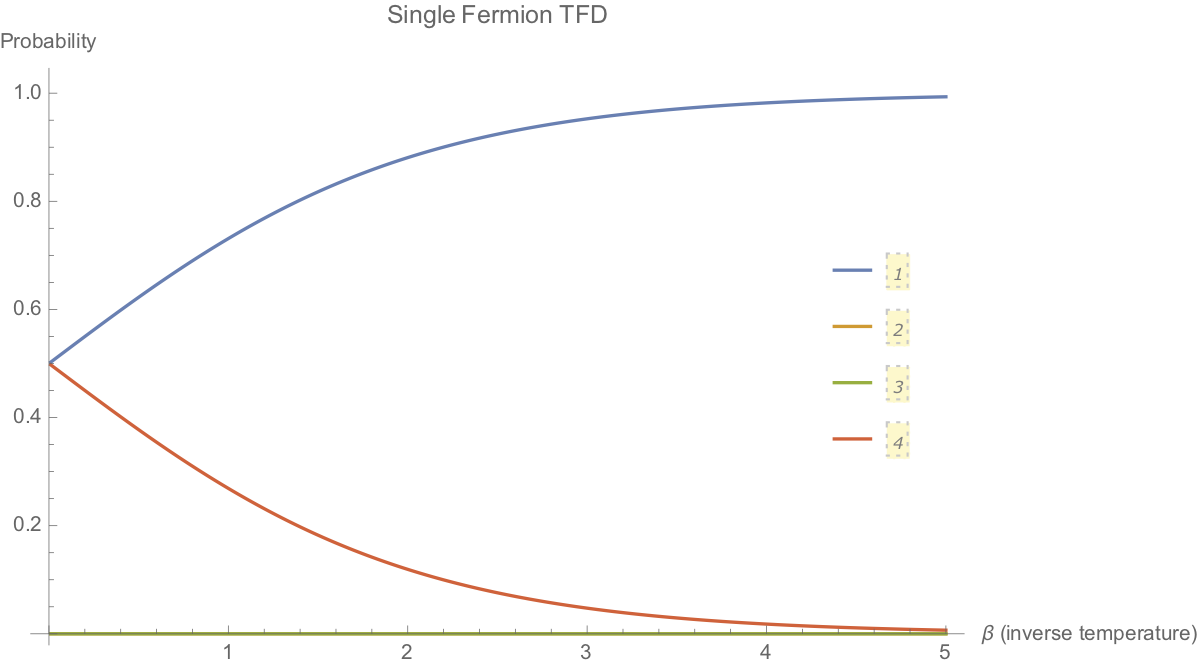}
    \caption{Plot of the temperature evolution of the state of a fermion and its thermo field double, calculated classically using Mathematica.}
    \label{fig:tfevo_m}
\end{figure}

\begin{figure}[H]
    \centering
    \includegraphics[width=0.8\linewidth]{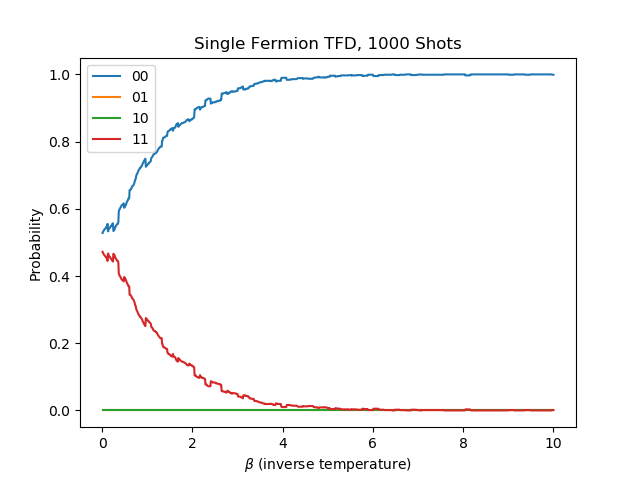}
    \caption{Plot of the temperature evolution of the state of a fermion and its thermo field double, calculated using Qiskit's quantum simulator.}
    \label{fig:tfevo_p}
\end{figure}

The results from Mathematica and Qiskit match up almost exactly. The roughness of the Qiskit plot are due to the discrete nature of the calculation: the measurement for each value of $\beta$ was repeated 1000 times, so what we're seeing is essentially 200 histograms in a single plot. If we increase the number of shots (measurements per step), the plot should become progressively smoother. The plot tells us that at high temperature (low $\beta$), the state of the qubits is a superposition of the $\ket{11}$ and $\ket{00}$ states. As the temperature decreases ($\beta$ increases), this mix decays exponentially until the qubits are in a pure $\ket{00}$ state. The mixed state $\ket{10}$ and $\ket{01}$ are never seen; this is because the thermo field double is always in the same state as the ``real'' fermion.

\section{Conclusion}

Due to the mathematical properties of the transformation we apply to our operators in thermo field dynamics, we cannot use the VQE algorithm in the usual way to find the ground state energy of modified Hamiltonians. Unless we find a way to modify the Hamiltonian which encodes information on temperature and changes its eigenvalues, the VQE algorithm won't be applicable here. In the future, we also plan to use techniques from  \cite{Cottrell:2018ash}, \cite{Wu:2018nrn}, \cite{Swingle:2016foj} and \cite{Maldacena:2018lmt} to perform a variational quantum computation of the thermo double state. Since the operator we use is similar to a time-evolution operator, we can use the techniques of quantum walks to analyze the effect of the operator on quantum states. The method we used relies on an algorithm in Qiskit which is currently limited to 2-qubit unitary matrices. Unless a similar algorithm is found for larger even numbers of qubits, we won't be able to perform similar calculation on larger systems with thermo field dynamics. We're still in the regime of small matrices, so these analyses can still comfortably be done classically for larger systems.

\section{Acknowledgements}

This project was supported in part by the Brookhaven National Laboratory (BNL), Computational Science Initiative under the BNL Supplemental Undergraduate Research Program (SURP). I would also like to thank my mentor Michael McGuigan for all of our stimulating discussions, as well as past and present CSI interns for their help setting up the Mathematica and Python code. We acknowledge use of the IBM Q for this work. The views expressed are those of the authors and do not reflect the official policy or position of IBM or the IBM Q team. Michael McGuigan is supported from DOE HEP Office of Science DE-SC0019139: Foundations of Quantum Computing for Gauge Theories and Quantum Gravity.

\bibliographystyle{unsrt}
\bibliography{references}

\begin{thebibliography}{10}

\bibitem{Das:1989cj}
Ashok~K. Das.
\newblock {Supersymmetry and Finite Temperature}.
\newblock {\em Physica}, A158:1--21, 1989.

\bibitem{Qiskit}
Gadi Aleksandrowicz, Thomas Alexander, et~al.
\newblock Qiskit: An open-source framework for quantum computing, 2019.

\bibitem{Peruzzo:2013nat}
A.~Peruzzo, J.~McClean, P.~Shadbolt, M.~Yung, X.~Zhou, P.~J. Love,
  A.~Aspuru-Guzik, and J.~L. O'Brien.
\newblock A variational eigenvalue solver on a quantum processor.
\newblock 2013.

\bibitem{Jagannathan:1981rh}
R.~Jagannathan, T.~S. Santhanam, and R.~Vasudevan.
\newblock {Finite Dimensional Quantum Mechanics of a Particle}.
\newblock {\em Int. J. Theor. Phys.}, 20:755, 1981.

\bibitem{Singh:2018qzk}
Ashmeet Singh and Sean~M. Carroll.
\newblock {Modeling Position and Momentum in Finite-Dimensional Hilbert Spaces
  via Generalized Clifford Algebra}.
\newblock 2018.

\bibitem{Atakishiyev:2008}
N.~M. {Atakishiyev}, A.~U. {Klimyk}, and K.~B. {Wolf}.
\newblock {A discrete quantum model of the harmonic oscillator}.
\newblock {\em Journal of Physics A Mathematical General}, 41(8):085201,
  February 2008.

\bibitem{Lorente:2001}
M.~Lorente.
\newblock Continuous vs. discrete models for the quantum harmonic oscillator
  and the hydrogen atom.
\newblock {\em Physics Letters A}, 285(3):119 -- 126, 2001.

\bibitem{Barker:2000}
Laurence Barker, Ãagatay Candan, Tugrul Hakioglu, M~Alper Kutay, and
  Haldun~M Ozaktas.
\newblock The discrete harmonic oscillator, harper's equation, and the discrete
  fractional fourier transform.
\newblock {\em Journal of Physics A: Mathematical and General}, 33(11):2209,
  2000.

\bibitem{Bogoljubov:1958}
N.~N. Bogoljubov.
\newblock On a new method in the theory of superconductivity.
\newblock {\em Il Nuovo Cimento (1955-1965)}, 7(6):794--805, Mar 1958.

\bibitem{Takahashi:1996zn}
Y.~Takahashi and H.~Umezawa.
\newblock {Thermo field dynamics}.
\newblock {\em Int. J. Mod. Phys.}, B10:1755--1805, 1996.

\bibitem{Cottrell:2018ash}
William Cottrell, Ben Freivogel, Diego~M. Hofman, and Sagar~F. Lokhande.
\newblock {How to Build the Thermofield Double State}.
\newblock 2018.

\bibitem{Wu:2018nrn}
Jingxiang Wu and Timothy~H. Hsieh.
\newblock {Variational Thermal Quantum Simulation via Thermofield Double
  States}.
\newblock 2018.

\bibitem{Miceli:2018}
R.~Miceli and M.~McGuigan.
\newblock Quantum computation and visualization of hamiltonians using discrete
  quantum mechanics and ibm qiskit.
\newblock In {\em Proceedings, 2018 New York Scientific Data Summit (NYSDS):
  Upton, USA, August 6-8, 2018}, pages 1--6, 2018.

\bibitem{Swingle:2016foj}
Brian Swingle and John McGreevy.
\newblock {Mixed s-sourcery: Building many-body states using bubbles of
  Nothing}.
\newblock {\em Phys. Rev.}, B94(15):155125, 2016.

\bibitem{Maldacena:2018lmt}
Juan Maldacena and Xiao-Liang Qi.
\newblock {Eternal traversable wormhole}.
\newblock 2018.

\end{thebibliography}

\end{document}